\documentclass[lettersize,journal]{IEEEtran}
\usepackage{amsmath,amsfonts}
\usepackage{array}
\usepackage[caption=false,font=normalsize,labelfont=sf,textfont=sf]{subfig}
\usepackage{textcomp}
\usepackage{stfloats}
\usepackage{url}
\usepackage{verbatim}
\usepackage{graphicx}
\usepackage{cite}

\usepackage{amsthm,amsopn}
\usepackage{bm}
\usepackage{hyperref}
\usepackage{url}
\usepackage[capitalise]{cleveref}
\usepackage{graphicx,caption,lipsum}
\usepackage{booktabs,multirow}
\usepackage{placeins}
\usepackage{algpseudocode,algorithm}
\usepackage{color}
\usepackage[table,xcdraw]{xcolor}
\usepackage{bbm}
\usepackage{enumerate}   
\usepackage{adjustbox}
\usepackage[normalem]{ulem}
\usepackage{fixltx2e}
\def\reg{{\rm\ooalign{\hfil
      \raise.07ex\hbox{\scriptsize R}\hfil\crcr\mathhexbox20D}}}
\usepackage{tikz}

\begin{document}

\title{C3-DINO: Joint Contrastive and Non-contrastive Self-Supervised Learning for Speaker Verification }

\author{Chunlei Zhang,~\IEEEmembership{Member,~IEEE}, Dong Yu,~\IEEEmembership{Fellow,~IEEE}
        % <-this % stops a space
\thanks{This paper was produced by the IEEE Publication Technology Group. They are in Piscataway, NJ.}% <-this % stops a space
\thanks{Manuscript received April 19, 2021; revised August 16, 2021.}}

\markboth{Journal of \LaTeX\ Class Files,~Vol.~14, No.~8, August~2021}%
{Shell \MakeLowercase{\textit{et al.}}: A Sample Article Using IEEEtran.cls for IEEE Journals}


\maketitle

\begin{abstract}
Self-supervised learning (SSL) has drawn an increased attention in the field of speech processing. Recent studies have demonstrated that contrastive learning is able to learn discriminative speaker embeddings in a self-supervised manner. However, base contrastive self-supervised learning (CSSL) assumes that the pairs generated from a view of $anchor$ instance and any view of other instances are all negative, which introduces many false negative pairs in constructing the loss function. The problem is referred as $class$-$collision$, which remains as one major issue that impedes the CSSL based speaker verification (SV) systems from achieving better performances. In the meanwhile, studies reveal that negative sample free SSL frameworks perform well in learning speaker or image representations. In this study, we investigate SSL techniques that lead to an improved SV performance. We first analyse the impact of false negative pairs in the CSSL systems. Then, a multi-stage Class-Collision Correction (C3) method is proposed (e.g., false negative pair filtering, InfoNCE re-weighting and ProtoNCE), which leads to the state-of-the-art (SOTA) CSSL based speaker embedding system. On the basis of the pretrained CSSL model, we further propose to employ a negative sample free SSL objective (i.e., DINO) to fine-tune the speaker embedding network. The resulting speaker embedding system (C3-DINO) achieves 2.5\% Equal Error Rate (EER) with a simple Cosine Distance Scoring (CDS) method on Voxceleb1 test set, which outperforms the previous SOTA SSL system (4.86\%) by a significant +45\% relative improvement. With speaker clustering and pseudo labeling on Voxceleb2 training set, a LDA/CDS back-end applying on the C3-DINO speaker embeddings is able to further push the EER to 2.2\%. Comprehensive experimental investigations of the Voxceleb benchmarks and our internal dataset demonstrate the effectiveness of our proposed methods, and the performance gap between the SSL SV and the supervised counterpart narrows further.             
\end{abstract}


\section{Introduction}
\IEEEPARstart{S}{peaker} verification (SV) aims to verifying whether an unknown speech utterance belongs to a claimed speaker identity. It serves as a natural and effective way for biometric identity authentication. The direct applications of SV can be audio forensics, computer access control, and telephone voice authentication (e.g., telephone banking). Besides that, speaker representation learned from a SV system also acts as a key component in downstream tasks such as, speaker diarization, text to speech synthesis, voice conversion, speaker adaptation of automatic speech recognition as well as targeted speech enhancement and separation systems~\cite{senior2014improving,ji2020speaker,jia2018transfer,zhang2020durian,wang2018voicefilter,wang2018speaker,zhang2021towards,zhang2022e2e}.

In the age of deep learning, tremendous progresses have been made in SV. One common paradigm for a deep learning based SV system is to learn a speaker discriminative network, and treat it as a speaker embedding extractor afterwards~\cite{snyder2017deep,zhang2017end,desplanques2020ecapa}. The state-of-the-art (SOTA) approaches for learning speaker representations usually rely on supervised deep neural networks and large labeled speaker corpora. Various neural network architectures, such as Time-Delay Neural Network (TDNN), ResNet and ECAPA-TDNN, were explored as the speaker embedding extractors~\cite{snyder2018x,hajavi2019deep,zhang2017end,zhang2018text,he2016deep,chung2020defence}. Effective loss functions, such as triplet loss, softmax loss and its large margin alternatives (e.g., angular softmax loss, additive margin softmax loss or large margin cosine loss etc.), were investigated to encourage more inter-speaker separation and intra-speaker compactness~\cite{liu2017sphereface, deng2019arcface,zhang2018text,chung2020defence}.
To aggregate the frame-level information to the utterance-level, different temporal pooling methods~\cite{snyder2018x,okabe2018attentive,zhang2017end2,xie2019utterance} were explored as an important step to normalize phonetic variability. To further improve the robustness and generalization for SV, noise and language robust models and training paradigms have been proposed and improve the system performance~\cite{yu2017adversarial,xia2019cross}.

Despite the fact that impressive progress has been made with fully supervised models, challenges still exist for practical SV systems. Fully supervised models are usually domain specific, in order to perform well in a particular domain, satisfying amount in-domain labeled data is essential~\cite{chung2018voxceleb2, zhang2019utd}. In many real cases, it is expensive and difficult to obtain sufficient data annotations. For example, when vast amount of service data is available to access, it is ideal to use those in-domain data to continue facilitating the service models. However, annotating the online data involves huge amount of efforts. And both machine and human annotators introduce their systematic errors/bias into the labels, which may prevent the supervised models from achieving the optimal performance\cite{hansen2015speaker}. 

Self-supervised learning (SSL) is a form of unsupervised leaning where data itself provides the supervision. SSL usually defines a proxy loss, which is directly or indirectly connected with the downstream tasks. In order to solve it, the SSL network is then forced to learn what we care about, e.g., speaker representation, content representation or semantic representation in speech, images and text~\cite{he2020momentum,chen2020simple,xia2021self,hsu2021hubert,devlin2018bert}. In this study, we focus the SSL for speech modality, but the methods examined here can be generalized in other domains. 

Depending on what signal resolution is employed to construct the proxy losses, the SSL in speech can be generally categorized into two directions: the frame-level approaches and the instance-level approaches. One popular realization of a frame-level SSL model is to reconstruct/predict masked targets, which is conditioned on the unmasked contextual signals~\cite{liu2020mockingjay,hsu2021hubert}. The learned intermediate representation are expected to be more informative by leveraging contextual content when a large amount of unlabeled speech data is available. In addition, recent studies show that better speech representations can be achieved by jointly modeling of generative/predictive and discriminative power of signal, which is usually implemented by introducing an regularization term or an inductive bias (e.g., enforcing codebook diversity in wav2vec 2.0 or denoising in WavLM)~\cite{oord2018representation,baevski2020wav2vec,chen2021wavlm}. Unlike the frame-level approaches, where the SSL features have to be derived from the proxy models to train the downstream tasks (usually with labels). The instance-level approaches are more directly linked to the final objective. The instance-level SSL models are formulated by distinguishing the input signal from their distractors (e.g., MoCo or SimCLR) or constructing a similarity measure within the different viewers of the input signal (e.g., BYOL or DINO)~\cite{he2020momentum,chen2020simple,grill2020bootstrap,caron2021emerging}.          

 As an emerging direction in speaker representation learning, SSL based models have been investigated in the literature~\cite{stafylakis2019self,sang2021self,ding2020learning,chen2021large,huh2020augmentation,inoue2020semi,xia2021self}. In~\cite{ding2020learning, chen2021large,chen2021wavlm}, SSL is employed as a pre-training method for subsequent supervised speaker embedding training, a better SV performance is achieved by leveraging a large amount of speech corpora. In~\cite{huh2020augmentation}, an augmentation adversarial training strategy is introduced to learn channel-invariant speaker representations, which enhances the robustness of learned CSSL models against extrinsic varibilities. In~\cite{inoue2020semi,xia2021self}, the authors extended the self-supervised framework to a semi-supervised paradigm where a small portion of the whole dataset is labeled, which provided a good direction to utilize the real-world data. In~\cite{sang2021self}, a self-supervised regularization term was proposed in addition to the siamese network, an improved SV performance was obtained in the Voxceleb benchmark. In~\cite{cho2021jhu}, A DINO based speaker embedding system demonstrates the potential of SSL models for SV, a 4.83\% EER on Voxceleb1 test set represents the current SOTA performance.   

In this work, we investigate to effectively utilize unlabeled speech data in the SSL framework for speaker representation learning. We develop class-collision correction ($\bf{C3}$) methods for CSSL based speaker embedding systems and jointly utilizing CSSL and non-CSSL training strategies to improve the SV performance. To do so, we first develop MoCo as our contrastive speaker representation learning backbone, which is built on our previously study~\cite{xia2021self}. Next, we explore different network architectures and find out the best configuration for speaker embedding training. Then, to alleviate the class-collision issue in the CSSL system, alternative ways of constructing the InfoNCE loss (i.e., false negative pair filtering, the InfoNCE loss re-weighting and ProtoNCE loss), which directs us to the best CSSL framework, i.e., the C3-MoCo model. Finally, inspired by the similarity/distribution matching based models (e.g., BYOL and DINO), where only positive pairs are constructed in the training process, we extend the DINO speaker embedding system via incorporating the CSSL models. We find that the C3-MoCo initialed DINO speaker embedding system (C3-DINO) significantly improves the system performance. Extensive experiments are conducted on the Voxceleb dataset and a Tencent's in-house SV dataset to justify the effectiveness of our proposed methods.         

Although more detailed explanations and analysis can be found throughout this paper, let us first summarize the core contributions here:
\begin{enumerate}

\item{From the system development viewpoint, we systematically investigate two effective SSL speaker embedding frameworks (i.e., MoCo and DINO) for SV. By leveraging different network architectures (i.e., TDNN, ResNetSE34L, and ECAPA-TDNN), the proposed systems outperform the SOTA SSL based results;} 

\item{
The impact of false negative pair in one batch is analysed w.r.t. different scales of training data. The analysis provides an insight of how would a CSSL model evolve in terms of different speaker numbers; }

\item{
For scenarios with limited training data, class-collision correction methods are proposed. Particularly, false negative pair filtering and InfoNCE loss re-weighting are examined, in combination with the ProtoNCE loss, we achieve a relative 8.6\% improvement over the SOTA contrastive learning based system~\cite{sang2021self}. }   	
      	
\item{
In addition, we introduce the C3-DINO speaker embedding framework---the CSSL initialed DINO speaker embedding system that incorporates both contrastive leanring and non-contrastive learning. We obtain a new SOTA SV performance with 2.5\% CDS EER in standard Voxceleb1 test benchmark, which represents a +48.6\% relative improvement over the previous best system. }

\item{
Together with clustering and pseudo labeling on the speaker embeddings of training data, a LDA/CDS back-end further narrows the performance gap between the SSL based system and the SOTA supervised models. 
}
\end{enumerate}
  	
The remainder of this paper is organized as follows. An overview of self-supervised learning frameworks and recent development for speaker recognition is presented in Sec. \ref{SSLbase}. In Sec.\ref{improved}, we first analyse the impact of class-collision to SV, and we propose methods that alleviate this issue. Then, we present the details of developing the C3-DINO speaker embedding system. The corpora and corresponding experimental setups for system development and evaluation are introduced in Sec.\ref{Corpora}. Sec.\ref{exps} details the experimental results. Finally, we conclude our work in Sec.\ref{cons}.

\section{Self-supervised speaker embedding baselines}
\label{SSLbase}

In this section, we first introduce the MoCo speaker embedding frame, where components such as momentum encoder, data augmentation are proved to be effective in most SSL frameworks. Inspired by the negative sample free SSL frameworks in computer vision and speech, we investigate a DINO based speaker embedding framework by changing only a small portion from the MoCo system. We analyse the similarity and difference between these two frameworks. At the end of this section, we present the implementation details of the proposed methods for developing the speaker recognition systems.

\begin{figure}[tbp]
    \centering
    \includegraphics[width=1\linewidth, height=5.7cm]{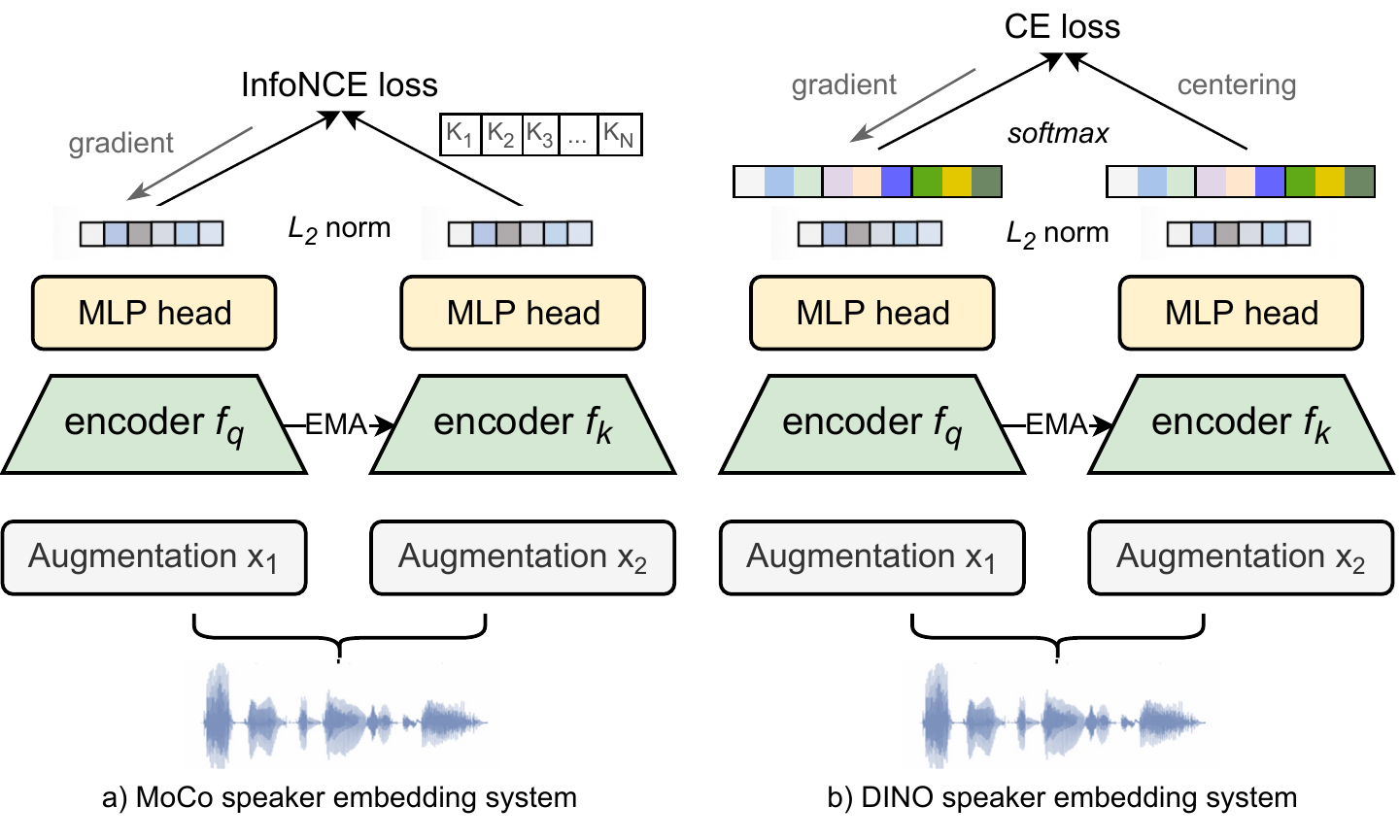}
    %\vspace{-ex}
    \caption{The baseline MoCo and DINO training diagrams.}
    \label{fig:se1}
    
\end{figure}

\subsection{MoCo speaker embedding system}

We start the MoCo framework description by introducing a simple contrastive learning approach (SimCLR)~\cite{chen2020simple}.  Both SimCLR and MoCo share a similar structure which comprises three major components:
(1) A data augmentation module that transforms a data example into two different views, which formulates a positive pair. (2) A neural encoder module that encodes the input features into a fixed latent space. (3) A MLP projection head that maps the latent representation to a $L_{2}$-normalized embedding (i.e., speaker embedding) where a contrastive loss is applied. 

SimCLR achieves the instance discrimination by directly maximizing the similarity between augmented positive pairs and minimizing the similarity of negative pairs via an InfoNCE loss~\cite{oord2018representation}.
For a mini-batch of $N$ samples, we can get $2N$ samples after two-fold augmentation. SimCLR doesn't sample negative examples explicitly. The rest $2(N-1)$ samples (different from the positive pair) are treated as negative examples without any filtering. The loss function is defined as,
\begin{align}
   \mathcal{L}_{SimCLR}=-\frac{1}{N}\sum_{i=1}^{N} \log \frac{\exp \left( \mathbf{v}_{i} \cdot \mathbf{v}^{\prime}_{i} / \tau\right)} 
   {\sum_{j=1}^{2N} \mathbbm{1}_{[i \neq j]} \exp \left( \mathbf{v}_{i} \cdot \mathbf{v}_{j} / \tau\right)},
\end{align}
where $ \mathbf{v}_{i}$ and $\mathbf{v}^{\prime}_{i}$ are the two augmented views of one sample, $ \mathbf{v}_{j}$ is traversed from the rest $2N$-1 augmented samples, $\mathbbm{1}_{[i \neq j]} $ is an indicator function that equals to 1 when $i \neq j$. $\tau$ is the temperature parameter that controls the sharpness of the output distribution, and $\cdot$ is simply the dot product of two $L_2$ normalized speaker embeddings.  

SimCLR uses samples in the current mini-batch for negative example mining. The size of negative pair is constrained with the mini-batch size $N$, which is limited by the GPU memory. To prevent this problem and introduce more negative samples in the InfoNCE loss, we also explored a momentum contrastive based speaker embedding system in our previous study~\cite{xia2021self}. We mainly followed the MoCo framework for visual representation learning~\cite{he2020momentum}. The MoCo speaker embedding system used a dynamic queue to store a large number of negative samples for calculating the loss, with an assumption that robust features can be learned by a large dictionary that covers a rich set of negative samples. As illustrated in the left side of \cref{fig:se1}, MoCo uses one query encoder $f_q$ to encode one copy of the augmented samples, and employs a momentum encoder $f_k$ to encode the other copy. The momentum encoder is a slowly progressing encoder that represents the snapshots of the previous query encoders. The InfoNCE loss employed in the MoCo speaker embedding learning is defined as,
\begin{align}\label{moco_eq}
\mathcal{L}_{MoCo} =-\frac{1}{N}\sum_{i=1}^{N}\log \frac{\exp \left(\mathbf v_{qi} \cdot \mathbf v^{\prime}_{{ki}} / \tau\right)}{\sum_{j=0}^{K} \exp \left(\mathbf v_{qi} \cdot \mathbf v_{kj} / \tau\right)},
\end{align}

\noindent 
where $\mathbf{v}_{qi}$ and $\mathbf v^{\prime}_{{ki}}$ is the query sample (i.e., $anchor$) and the corresponding key ($positive$ sample) in the mini-batch. The positive sample $\mathbf v^{\prime}_{{ki}}$ is encoded with the momentum encoder $f_k$, while the $negative$ samples $\mathbf v_{{kj}} $ in the denominator are the inventory key speaker embeddings in the queue. The $\mathcal{L}_{MoCo}$ in \cref{moco_eq} is the average negative log score ratios over all the $N$ positive pairs and the corresponding $K$ $negative$ pairs. The introduction of a queue with the size $K$ is able to enlarge the number of negative samples without consuming two much memory. At the same time, samples in the dictionary are progressively replaced, with the latest mini-batch of key embeddings popped-in and the oldest popped-out. $K$ was set to 10000 in our previous experiments, we found that MoCo was not sensitive to $K$ after a certain value.

To make the key embeddings in the negative sample queue consistent, the key encoder $f_k$ is updated as an Exponential Moving Average (EMA) of the query encoder $f_q$. Denoting the parameters of $f_k$ as $\theta_k$ and those of $f_q$ as $\theta_q$, we update $\theta_k$ by $\theta_{\mathrm{k}} \leftarrow m \theta_{\mathrm{k}}+(1-m) \theta_{\mathrm{q}}$, where $m$ = 0.996 is the momentum coefficient. Only the parameters $\theta_q$ are updated by back-propagation during training. 
The momentum update makes the key encoder evolve more smoothly than the query encoder. 
With this design, though the keys in the queue are encoded by different encoders, the difference among these encoders is small. Therefore we can maintain a sizeable negative sample queue in a stable training process. In our study \cite{xia2021self}, we found the MoCo based speaker embedding achieved a considerable performance gain over the SimCLR system. Thus, MoCo is considered as the first baseline in this study. 

\subsection{DINO speaker embedding system}
\label{dino-baseline}
The MoCo based speaker embedding systems compose the negative samples that are different from the $anchor$ sample to make the anchor and negative samples far from each other in the embedding space. However, this could be problematic as the negative sample queue accumulates false negative samples, i.e., the anchor and negative utterances may share the same speaker identity. This is referred as the $class$ $collision$ problem in the CSSL frameworks, which results in the major performance deficit compared with the supervised contrastive learning systems~\cite{khosla2020supervised}. On the other hand, non-contrastive methods, such as BYOL and DINO, have demonstrated competitive performance compared with the constrastive methods in image or speech modality~\cite{grill2020bootstrap,caron2021emerging,cho2021jhu}. Instead of discriminating between the instances, both BYOL and DINO embrace the knowledge distillation methodology in their system design, BYOL operates at the embedding level and DINO distills the knowledge from the momentum encoder (teacher) with an additional softmax layer. In this study, we develop a DINO speaker embedding system as the right hand side of \cref{fig:se1}, which requires us the minimum changes from the MoCo baseline system.

%\subsubsection{Self-distillation with no labels}
Similar to MoCo, the base DINO system requires two differently augmented samples as the input to the student branch (a local view to query encoder $f_{q}$) and the teacher branch (a global view key to encoder $f_{k}$), and $f_k$ is also updated as the EMA of the $f_q$. More specifically, we keep every components the same as the baseline MoCo until the speaker embedding layer, where a softmax layer is added after the embedding layer in each branch. Following~\cite{caron2021emerging}, we minimize the cross-entropy between two distributions calculated along the two branches, the DINO loss is simply represented as,
\begin{align}\label{CE1}
\mathcal{H}(X,Y) =-\sum_{i=1}^{K}p((x_{i}-c_{i})/\tau_{t})\log q(y_i/\tau_{s}), 
\end{align}
where $p((x_{i}-c_{i})/\tau_{t})$ represents the probability of class (position) $i$ after the teacher branch mapping, $x_{i}-c_{i}$ is the subtraction between $i^{th}$ logit of the global view $X$ and the center $C$, denoted as $C = mC+ (1-m)\frac{1}{B}\sum_{i=1}^{B} X^i $, representing the EMA of the batch centers ($m=0.99$). $\tau_{t}$ is the teacher branch temperature coefficient. Similarly, $q(y_i/\tau_{s})$ is the probability of class (position) $i$ after the student branch mapping and student temperature normalization, $y_i$ is the $i^{th}$ logit of the local view $Y$, $K$ is the total number of classes assigned in the DINO training. The choice of $\tau_{t}$ and $\tau_{s}$ determines the direction of the self-distillation process. A relative smaller value of $\tau_{t}$ (comparing to $\tau_{s}$) sharps the teacher distribution and forces the distribution of the local view in the student branch to be close to the teacher distribution, such that information concentration happens. The cross-entropy without the temperature scaling $\mathcal{H}(p,q) =-\sum_{i=1}^{K}p(x_{i}-c_{i})\log q(y_i)$ ends up with a flat distribution across the whole data set. And the cross-entropy $\mathcal{H}(p,q)= -\sum_{i=1}^{K}p(x_{i})\log q(y_i)$ without centering results in a trivial solution that one position dominates for all samples. 

Ideally, a good representation learning should map both local and global sample of the sample speaker to the same the probabilistic pattern with a peaky distribution in one position. Meanwhile, samples from different speakers should be allocated on different positions. Since center $C$ is the EMA of batch centers, we argue that the converged logits of center $C$ should be flatten across all $K$ positions. The probability $p((x_{i}-c_{i})/\tau_{t})$ can be approximately written as,
\begin{align}\label{softmax1}
p((x_{i}-c_{i})/\tau_{t})  & = \frac{\exp^{(x_{i}-c_{i})/\tau_{t}}}{\sum_{i=1}^{K} \exp^{(x_{i}-c_{i})/\tau_{t}} }, \\
 & \approx \frac{\exp^{x_{i}/\tau_{t}}}{\sum_{i=1}^{K} \exp^{x_{i}/\tau_{t}} } ,
\end{align}
This approximation is only hold when $c_i$ in center $C$ follows a uniform distribution, and the DINO loss of \cref{CE1} is converged to a standard cross-entropy form with instance discrimination ability. Since the center $C$ is the EMA of the batch centers, to satisfy the uniform distribution assumption, we can either choose to use a big batchsize or employ a small update parameter $1-m$, so that the training is stable. At the same time, we argue that K should be {\bf{enough big}}, so that the training samples with different underlying speaker identities will be allocated to appropriate locations without the confusion. After all these, the interpretation of the converged DINO training is that: the similarity between the anchor-positive pair is maximized, while center $C$ is located in the middle. DINO is explored as our second baseline speaker embedding system.

\subsection{Implementation details} 
\noindent\textbf{Feature extraction:} we compute 30-dimensional MFCC on the frame level as the input features. A Hamming window of length 25 ms with a 10 ms frame-shift is used to extract the MFCC from audio signals. Similar to \cite{xia2021self}, we use a random chunk of 200-400 frame features of each audio file as the input to the MoCo systems. While for DINO experiments, we randomly use 300-400 frames as the global view to $f_k$, and chunk 50\% length of the corresponding global view as the local view to $f_q$. Since no zero padding is applied to the feature processing, we filter out the utterances that are shorter than 400 frames. Each input feature is mean normalized on the frame level. An energy-based VAD is used to remove silent frames, which follows the Kaldi SRE16 recipes~\cite{povey2011kaldi}. We sample 600,000,000 frames for each epoch of training in our following experiments.  

\noindent\textbf{WavAug:} follow the conclusion from~\cite{xia2021self}, we only perform data augmentation at the wavform level. Different from the recipes for supervised models, where only one kind of extrinsic variabilities is applied to the raw sample. We find that \textit{progressive} data augmentation is more effective for both MoCo and DINO systems. Specifically, we first augment one utterance to two views with reverberation by randomly selecting RIRs in the OpenSLR Room Impulse Response and Noise Database~\cite{ko2017study}, we pool all the RIRs from different size of rooms together in this step. There is a 80\% probability that the sample is reverb-augmented, while the rest 20\% of the total utterances are kept unchanged. After the reverb-augmentation, we further add one of the three kinds of noises with equal probability: 1) Noise, 2) Music, and 3) Babble. In each augmentation, one noise file is randomly selected in the MUSAN database~\cite{snyder2015musan} and added to the recording with [0, 5, 10, 15] dB SNR. Alternatively, one music file is randomly selected and added to the recording with [5, 8, 10, 15] dB SNR. Otherwise, one human babble speech is selected and added to the recording with [13, 15, 17, 20] dB SNR. We repeat the \textit{progressive} data augmentation process with a different random seed to incorporate enough extrinsic variabilities.

\noindent\textbf{Network architectures:} three different networks are employed to examine their impact to the SV performance. The first one is a standard TDNN model, it has five 1-D convolutional layers and uses a statistical pooling layer to encode the variable-length feature input to a fixed 3000 dimension vector. The MLP projection head contains two layers. The dimentionality of the first MLP layer is 512, followed by a batch normalization layer and ReLU activation. The second MLP is a 128-D linear transformation with a $L_2$ normalization applied, where it is used as the speaker embedding layer. The second network architecture in our study is a light version of ResNet34 (ResNetSE34L)~\cite{chung2020defence},  with the kernel size of the first convolution layer as 3 instead of 7. After a sequence of convolutional layers and residual connections, a statistical pooling layer is added to aggregate the frame-level features to the utterance-level. Like the TDNN implementation, the same MLP layers is applied to the end of ResNetSE34L. The last encoder architecture is the ECAPA-TDNN, which represents
the SOTA performances in the Voxceleb benchmark in supervised learning~\cite{desplanques2020ecapa}. In the ECAPA-TDNN, we only use one 128-D linear layer with batch normalization and $L_2$ normalization. For DINO implementation, we add a $K$ dimensional softmax layer after the speaker embedding layer, so that the cross-entropy loss is calculated. For system training, we use a ADAM optimizer and initialize the learning rate with $1e$-3. The learning rate is reduced to the stopping learning rate $10^{-5}$ with a cosine learning rate scheduler. The parameters of each kind of network is listed in \cref{model_size}.

\begin{table}[!t]
\caption{Encoder sizes w.r.t. three main stream architectures.}
\label{model_size}
\centering

\begin{tabular}{|c||c|c|c|}
\hline
name & TDNN & ResNetSE34L & ECAPA-TDNN \\
\hline
size (M)& 5.2 & 1.4 & 14.1\\
\hline
\end{tabular}
\end{table}

\noindent\textbf{Back-end scoring method and evaluation metric:} a simple CDS is used as the main scoring function since the setup for SV in this study is unsupervised. We also explore a self-training strategy (clustering with initial SSL speaker embeddings and assigning pseudo labels) at back-end level, where a conventional LDA/CDS classifier is utilized. The performance is reported in terms of Equal Error Rate (EER) and minimum Detection Cost Function (minDCF) with $P_{target} = 0.01$. 

\section{Improving the SSL based speaker embedding systems}
\label{improved}
In order to further improve the baselines, we firstly analyse the impact of false negative pairs to SV performance in the MoCo framework. Then, we propose a progressive class-collision correction (C3) method that boosts the performances. Finally, we propose the C3-DINO system, a framework that incorporates both contrastive learning and non-contrastive learning for speaker embedding learning, which leads to the SOTA SSL based SV system. 
\subsection{The impact of false negative pairs in MoCo systems }
\label{pfn_analysis}
In general, we know that the false negative pairs contradict with the ideal InfoNCE loss design, however there is no study that explicitly analyses how false negative pair impacts the overall SV performance. With supervision, we can find how many false negative pairs are actually in the denominator ${\sum_{j=0}^{K} \exp \left(\mathbf v_{qi} \cdot \mathbf v_{kj} / \tau\right)}$ of \cref{moco_eq}. In this portion of study, we consider the training strategy like MoCo-v3~\cite{chen2021empirical} for simplicity, where a mini-batch with $N$ samples is used for constructing the negative pairs (instead of a large queue with $K$ samples). Here, we use $p_{fn}$ to define the probability of having at least one false negative pair in the denominator of contrastive loss component, which is represented as:
\begin{align}\label{pfn}
p_{fn} = \frac{1}{N} \sum_{i=1}^{N} \mathbbm{1}_{[y_{i} = y_{j}  \bigcap j\neq i]}, j\in [1,N]
\end{align}
where $y_{i}$ is the label of the query embedding $\mathbf{v}_{qi}$, $y_{j}$ is the label of the key embedding $\mathbf{v}_{kj}$.  $\mathbbm{1}_{[y_{i} = y_{j} \bigcap j\neq i]},j\in [1,N] $ indicates the situation when $\mathbf{v}_{qi}$ is given as the anchor sample and there is at least one false negative pair in the contrastive loss component $\mathcal{L}_{i} = -\log \frac{\exp \left(\mathbf v_{qi} \cdot \mathbf v^{\prime}_{{ki}} / \tau\right)}{\sum_{j=1}^{N} \exp \left(\mathbf v_{qi} \cdot \mathbf v_{kj} / \tau\right)}$. With \cref{pfn}, we can conduct a series of experiments with controlled $p_{fn}$, so that we can a clear understanding of its impact to MoCo based SV systems.

We conduct experiments with Voxceleb 1 and Voxceleb 2 training data to examine the relations between $p_{fn}$ and data size, the results on Voxceleb 1 test set is reported. A MoCo baseline with the TDNN encoder as the backbone is used, the batchsize $N$ is set to 4096. In \cref{pfn_eer}, a comprehensive analysis on the impact of $p_{fn}$ to EER is illustrated. It is easy to notice that the starting $p_{fn}$ for Voxceleb1 experiment is much higher than the Voxceleb2 MoCo $p_{fn}$, which shows that a bigger dataset with more speakers is very important in CSSL frameworks. The $p_{fn}$ ranges from fully self-supervised MoCo with the highest $p_{fn}$ to fully supervised contrastive loss with $p_{fn}=0$~\cite{khosla2020supervised}. To implement $p_{nf}$ with a controlled rate, we just randomly drop out the contrastive loss component $\mathcal{L}_{i} = -\log \frac{\exp \left(\mathbf v_{qi} \cdot \mathbf v^{\prime}_{{ki}} / \tau\right)}{\sum_{j=1}^{N} \exp \left(\mathbf v_{qi} \cdot \mathbf v_{kj} / \tau\right)}$ that contains false negative pairs.

\begin{table}[!h]
\caption{Performance analysis (EER, in \%) w.r.t. different $p_{fn}$ (in \%) in the supervision signal controlled contrastive learning systems.}
\label{pfn_eer}
\centering

\begin{tabular}{|c|c||c|c|c|c|c|c|}
\hline
Vox & $p_{fn}$ & 94.6(MoCo) & 80 & 60 & 40 & 20 & 0 \\

celeb1 &EER & 11.3 & 10.8 & 10.3 & 10.0& 9.4 & 7.3 \\
\hline
\hline
Vox & $p_{fn}$ & 61.2(MoCo) & 50 & 40 & 30 & 20 & 0 \\
celeb2 &EER & 8.5 & 7.9 & 7.8 & 7.6& 7.3 & 5.7 \\
\hline
\end{tabular}
\end{table}

As we can see from \cref{pfn_eer}, the SV performance improves with less false negative pairs for both datasets. The performance change seems not to be linearly aligned with the differences between different $p_{fn}$. For example, the EER improvement is relative less when $p_{fn}$ is large. And there is a big difference in the SV performance when $p_{fn}$ changes from 20\% to 0\%, which indicates that even a small portion of false negative pairs lead to an obvious performance degradation. That is the major reason that impedes the CSSL algorithms from achieving a better system performance.

\begin{figure*}[t]
    \centering
    \includegraphics[width=0.95\linewidth, height=5cm]{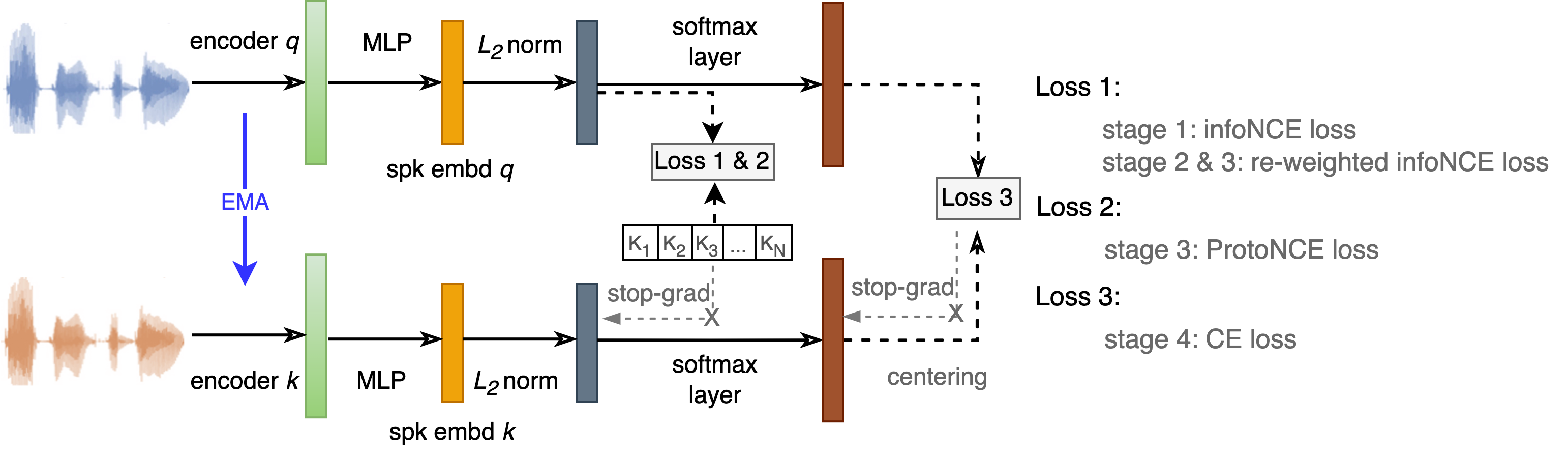}
    \caption{A multi-stage training framework. For the C3-MoCo training, the first stage is trained the infoNCE loss, then replace the InfoNCE loss with the re-weighted InfoNCE loss for the second stage, and jointly train with the ProtoNCE loss for stage 3. For C3-DINO training, the best way is using the pretrained C3-MoCo model as the initial, and apply the DINO loss.}
    \label{fig:se2}
\end{figure*}

\subsection{C3-MoCo speaker embedding systems }
\label{C3}
Based on the analysis in \cref{pfn_analysis}, we propose class-collision correction methods that can effectively alleviate the impact of false negative samples. An intuitive way is to find out the contrastive loss components $\mathcal{L}_{i}$ that contains false negative pairs, and filter them out in the InfoNCE loss calculation. Since label is not available, it is difficult to perfectly separate them in practice. Thus, we design a {\bf{false negative pair filtering}} strategy based on the cosine similarity of the anchor-positive pair ($\mathbf v_{qi} \cdot \mathbf v^{\prime}_{{ki}}$) and anchor-negative pairs ($\mathbf v_{qi} \cdot \mathbf v_{{kj}}, j\in [1,K]$), $K$ is the queue size in \cref{moco_eq}. Specifically, when $\mathcal{L}_{i}$ satisfies both conditions as,    
\begin{subequations}\label{eq:cond}
\begin{align}
\mathbf v_{qi} \cdot \mathbf v_{{kj}} &>  0.8 \times \mathbf v_{qi} \cdot \mathbf v^{\prime}_{{ki}}, \space j\in[1,K], \label{eq:condA}\\
\mathbf v_{qi} \cdot \mathbf v^{\prime}_{{ki}} &> 0.4, \label{eq:condB}
\end{align}
\end{subequations}
\cref{eq:condA} means if there are negative pairs that have a high cosine similarity (larger than 80\% of anchor-positive similarity), at the same time, \cref{eq:condB} requires that similarity between anchor-positive samples should be at a certain value, then $\mathcal{L}_{i}, i\in[1,N]$ ($N$ is the batchsize) is considered a false negative term. The purpose of \cref{eq:condB} is to keep $\mathcal{L}_{i}$ when the anchor-positive pair is not close enough (0.4 in our setup), and address them specifically during the training. It is noted that the speaker embeddings should be discriminative for the above operation, which is the prerequisite of this filtering strategy. 

In practice, just using a hard criterion to filter out samples has two drawbacks: a) the samples satisfied with \cref{eq:cond} may still contain false negative components; b) \cref{eq:cond} filters out the hard samples, which may be useful to further improve the system. So, we employ a {\bf{re-weighted InfoNCE loss}} to replace the MoCo loss in \cref{moco_eq},
\begin{align}\label{reweight_eq}
\mathcal{L}_{re\_MoCo} = \omega_1 \mathcal{L}_{MoCo}^{tn} + \omega_2 \mathcal{L}_{MoCo}^{fn},
\end{align}
where $\mathcal{L}_{MoCo}^{tn}$ is the MoCo loss components with the predicted true negative pairs, while $\mathcal{L}_{MoCo}^{fn}$ is representing the MoCo loss components with the predicted false negative pairs. And the $\mathcal{L}_{MoCo}^{fn}$ corresponds to elements with at least one false negative (predicted) in the key set. For each $\mathcal{L}_{i}$, the denominator is always different, since the anchor $\mathbf v_{qi}$ is different. We split the $\mathcal{L}_{i}$ across $N$ samples (a batch) into a ``TN" set and a ``FN" set, calculate the average MoCo losses ($\mathcal{L}_{MoCo}^{tn}$ and $\mathcal{L}_{MoCo}^{fn}$), respectively. Finally, we use $\omega_1$ and $\omega_2$ in Eq. \ref{reweight_eq} to re-weight the two MoCo losses. In summary, we filter the MoCo loss component $\mathcal{L}_{i}$ into the {\it predicted} true negative set and the false negative set, and adjust the weight to alleviate the class-collision issue. In our experiments on the Voxceleb benchmark, we find that $\omega_{1} = 0.8 $ and $\omega_{2} = 0.2 $ is a good configuration that encourages more contributions from $\mathcal{L}_{MoCo}^{tn}$ and still keeps a small portion of information from the hard samples. 

The last step we do to improve the MoCo baselines is the employment of the {\bf{ProtoNCE loss}}~\cite{xia2021self}. This is done after the re-weighted InfoNCE loss training is converged. Similar to~\cite{xia2021self}, a prototype memory bank is used to store the cluster centriods, the centriod is derived via a k-means clustering on the whole training dataset. The overall objective function $\mathcal{L}_{C3\_MoCo}$ is formulated as, 
\begin{align}\label{c3moco}
&\mathcal{L}_{C3\_MoCo} = \mathcal{L}_{re\_MoCo}  
+ \alpha \mathcal{L}_{pNCE}, \\
&\mathcal{L}_{pNCE} =-\frac{1}{N}\sum_{i=1}^{N} \log \frac{\exp \left(\mathbf{v}_{qi} \cdot \mathbf{c}_{s} / \phi_{s}\right)}
{\sum_{j=0}^{R} \exp \left(\mathbf{v}_{qi} \cdot \mathbf{c}_{j} / \phi_{j}\right)},
\end{align}
where $\mathcal{L}_{pNCE}$ is the ProtoNCE loss, $\mathcal{L}_{re\_MoCo}$ is used to retain the property of local smoothness and help bootstrap clustering. $\alpha$ is a weight coefficient to balance two losses, it is set to 0.2 in our experiments. For an anchor example $\mathbf{v}_{qi}$ (the query embedding), the positive embedding $\mathbf{c}_s$ is the prototype of class $s$ (i.e., the centroid of cluster $s$ where $\mathbf{v}_{qi}$ is assigned). Different class prototypes $\{\mathbf{c}_j\}_{j=1}^{R}$ are sampled as the negative examples, $R$ is the number of negative pairs, $R=10000$ in our experiments. Following the studies in~\cite{li2020prototypical,xia2021self}, we also the dynamically estimated temperature coefficient, $\phi=\frac{\sum_{i=1}^{Z}\left\|\mathbf{v}_{i} - \mathbf{c} \right\|_{2}}{Z \log (Z+\epsilon)}$, which indicates the concentration level of embeddings $\{\mathbf{v}_i\}_{i=1}^{Z}$ around their class centroid $\mathbf c$, $Z$ is total number of embeddings in that class and $\epsilon$ is a small constant for numerical stability. 

In \cref{fig:se2}, we have illustrated the progressive multi-stage stage training strategy for training the C3-MoCo speaker embedding system. The corresponding experimental results are presented in \cref{exps}.

\subsection{C3-DINO speaker embedding system}
\label{C3-DINO}
In this section, we explore methods that further improve the SSL speaker embedding system. There exists two ways to fully utilize the MoCo loss and the non-contrastive DINO loss. The first one is the joint training with a multi-task objective, 
\begin{align}\label{c3dino}
&\mathcal{L}_{C3\_DINO} = \mathcal{L}_{C3\_MoCo}  
+ \beta \mathcal{L}_{DINO}, 
\end{align}
where $\beta$ is the weight to balance the contrastive loss and the DINO loss (cross-entropy). However, one concern is that the individual loss may be converged to different levels, especially when there is performance gap between the single objectives, then the multi-task loss will not benefit much to the final results. We refer the multi-task training based system as C3-DINO$_1$.

System pretraining is the second way to incorporate both MoCo and DINO speaker embedding systems during training. Since the encoder design for both frameworks is the same, the whole C3-DINO speaker embedding system can be unified as \cref{fig:se2}. We firstly have the C3-MoCo system trained, and then use a single DINO loss to fine-tune the pretrained model. It is noted that an additional softmax layer is added in the DINO system, so we restore the learning rate to $1e$-3 and train the system until convergence. We use C3-DINO$_2$ to denote the C3-MoCo initialed DINO system.         
\section{SSL Corpora}
\label{Corpora}
Two corpora are utilized in the experimental studies. The first one is the Voxceleb corpus, which is our primary benchmark dataset to evaluate the proposed systems. The second one is our internal SV dataset, the test set is collected under the unconstrained vehicle environments in Mandarin (noted as ``vehicle-spk")~\cite{zhang2021towards}, the details of the corpora are listed below. 

\noindent{\textbf{Voxceleb:}} we use Voxceleb1 and Voxceleb2 training data for training the models. The Voxceleb1 training set contains 1211 speakers with 148642 utterances in total. With two-fold WavAug, we have 297284 utterances for each view. The Voxceleb2 training set has 5994 speakers, the total utterance number is 1092009. Similarly, we produce 2184018 utterances for one sample view (4368036 with two sample views combined). We use offline augmentation, we find a significant training speedup when we use the offline feature. For Voxceleb experiments, two test sets are used. The first one is the standard Voxceleb1 test set, which is the most popular benchmark recently. The other one is the VoxSRC21 dev set, it is drawn from the Voxceleb1 training set and introduced in VoxSRC 2021 challenge. 

\noindent{\textbf{Vehicle-spk:}}
the vehicle-spk set contains of 12000 speaker drawn from two gender-balanced public speech recognition datasets\footnote{http://en.speechocean.com/datacenter/details/254.htm} (8800 speakers) and one private SV corpus (3200 speakers) collected at Tencent. The total utterance number of the training set is 3.2M. The test set is a 365-speaker dataset collected under the unconstrained vehicle environments. Each speaker has 40-95 utterances, with a average number as 60. It is a comprehensive dataset that covers most of the noise types (i.e., wind noise, mechanical noise, background music/noise, and interference speakers etc.). 100K trials are produced in the end, among which 10K trials are target trials. The purpose of this portion of experiment is to study the behavior under different sizes of training data. 
\section{Experiments}
\label{exps}
In this section, we present detailed results with corresponding experimental configurations.
\subsection{Improving the MoCo and DINO baselines}
Network architecture is one of the most important driven forces in modern deep learning based SV systems. Although we focus more on the methodological side of training a SSL based speaker embedding model, we explore three milestone architectures to learn how each model behaviors~\cite{snyder2015time,he2016deep,desplanques2020ecapa}. We also discuss hyper-parameters that are important to the proposed MoCo and DINO systems.    

\noindent{\textbf{Comparing across network architectures}}: in \cref{Moco_baseline_results}, we list the SV performances w.r.t. three encoder types. Since the size of each neural network architecture is different, we set different batchsizes for TDNN (4096), ResNetSE34L(4096) and ECAPA-TDNN(1024) for MoCo experiments. In the DINO experiments, the batchsize is 1600 for both TDNN and R34L, 400 for ECAPA-TDNN. We find that MoCo and DINO is not much sensitive to batchsize with such configurations. It is noted that we use the $multi$-$crop$ trick for the DINO training~\cite{caron2021emerging}, specifically, 2 global views and 4 local views are generated from a pair of samples. That is the reason why significantly smaller batchsize is used in the DINO system. The training is conducted on a 8-GPU (P40) machine with 24GB RAM for each card.  
 
\begin{table}[htbp]
\caption{The SV performances of MoCo and DINO baseline systems w.r.t. three main stream architectures on Voxceleb1 test set. We use R34L and E-TDNN to represent ResNetSE34L and ECAPA-TDNN to save some space. }
\label{Moco_baseline_results}
\centering
\begin{tabular}{c|c|c@{\hskip 0.06in}c@{\hskip 0.06in}c|c@{\hskip 0.06in}c@{\hskip 0.06in}c}
\toprule
  Training & &\multicolumn{3}{c}{\underline{\textbf{EER}}(in\%)} & \multicolumn{3}{c}{\underline{\textbf{minDCF}}}  \\
% \cmidrule(r){4-6} \cmidrule(r){7-7} \cmidrule(r){8-8}
 Data & & TDNN & R34L & E-TDNN & TDNN & R34L & E-TDNN \\
\midrule
 Vox1 & MoCo& 11.3 & 10.8 & \textbf{9.8} & 0.76 & 0.72 & \textbf{0.67}\\
 Vox2 &  MoCo & 8.5 & 8.3 & \textbf{7.3} & 0.63 & 0.62  & \textbf{0.61}\\
 \midrule
 Vox1 & DINO & 7.2 & 6.8 & \textbf{6.1} & 0.76 & 0.72 & \textbf{0.52}\\
 Vox2 & DINO & 5.6 & 4.6 & \textbf{4.0} & 0.51 & 0.50 & \textbf{0.48} \\
\bottomrule
\end{tabular}
\end{table}
From \cref{Moco_baseline_results}, we can see that ECAPA-TDNN outperforms both TDNN and ResNetSE34L with a big margin, showing the advantages of ECAPA-TDNN in SSL based speaker embedding learning even without a large batchsize. For the MoCo baseline, the 7.3\% EER reproduces the result of~\cite{thienpondt2020idlab} and outperforms our previous TDNN based MoCo systems~\cite{xia2021self}. For DINO development, we set the softmax layer dimension $K$=$10000$ as the initial starting point, and report the results based on that. For all three models, our systems have performance gains over the previous SOTA DINO speaker embedding system~\cite{cho2021jhu}. The difference between the best MoCo and DINO baseline is surprisingly big, which may indicate that class-collision issue is severe, especially in the benchmark evaluation with limited training data. One may notice that the DINO system outperforms the SupCon system with the same TDNN encoder. It is because, we filter out $\mathcal{L}_i$ that contains false negative pairs, which largely reduces the training samples in each batch. At the same time, the result shows the advantages of the non-contastive loss in representation learning.        

\noindent{\textbf{Dimension of the softmax layer in DINO systems}}: in \cref{dino-baseline}, we have analysed that the dimension of the softmax layer $K$ is a critical hyper-parameter in the DINO system. To justify how the size $K$ impacts the system, we conducted a series of experiments where $K$ is the only variable. We draw a curve that based on EERs with different $K$ values . 

\begin{figure}[hbtp]
    \centering
    \includegraphics[width=0.85\linewidth, trim=4.5cm 9cm 4.5cm 8.7cm, height=5.7cm]{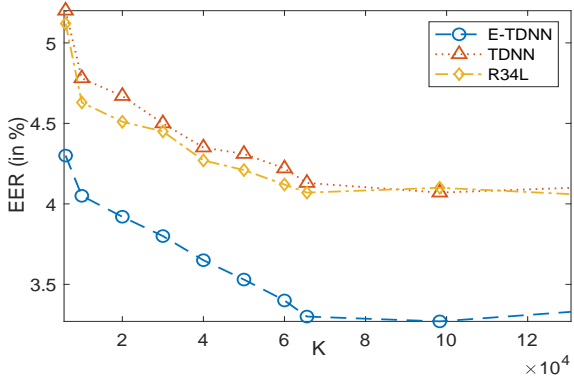}
    %\vspace{-ex}
    \caption{The SV performance (EER) of the DINO speaker embedding systems w.r.t. different dimension $K$. Specifically, $K$ is [6000, 10000, 20000, 30000, 40000, 50000, 60000, 65536, 98304, 131072].}
    \label{fig:k}
    
\end{figure}

As shown in \cref{fig:k}, all three networks share a similar trend, which achieves a fast improvement when $K$ is small. Then, the improvement is almost linearly aligned $K$ in a certain range and becomes flatten after that range. We find $K$=65536 is a good configuration that balances the batchsize (memory) and performance, achieving EER=3.3\% on Vox1 test set with Voxceleb 2 as the training data. In the experiments that follows, we fix $K$ as 65536.     

\subsection{C3-MoCo speaker embedding system for Voxceleb benchmarks}
As the described in \cref{C3}, we improve the baseline MoCo speaker embedding training via progressively introducing the re-weighted infoNCE loss and the ProtoNCE loss. We refer the final MoCo system as C3-MoCo. In our implementation, we first run the baseline MoCo training for 80 epochs, then we use the re-weighted infoNCE loss to replace \cref{moco_eq} for another 20 epochs, finally we use the $\mathcal{L}_{C3\_MoCo}$ of \cref{c3moco} to continue the training for the last 30 epochs. As illustrated in \cref{c3_eer}, the C3-MoCo system achieves 12.3\% relative improvement over the baseline MoCo system, which justifies that coping with the class-collision issue is important.

\subsection{C3-DINO speaker embedding system for Voxceleb benchmarks}
In the \cref{C3-DINO}, we describe two methods that incorporate both the contrastive and the non-contrastive loss for training. \cref{c3_eer} demonstrates the C3-DINO system with different training strategies. As we can see, the C3-DINO$_1$ system with a mutli-task loss does not outperforms the MoCo initialled DINO systems (i.e., C3-DINO$_2$). It is mainly because that there is a nontrivial performance gap between the single-loss MoCo and DINO systems, thus the training with \cref{c3dino} converges to a point in the middle. For C3-DINO$_2$ systems, we can observe that a better initial model results in a better final model. With 2.5\% EER is the best plain SSL based outcome, which represents a 48.6\% relative improvement over the SOTA SSL based system~\cite{cho2021jhu}. It is interesting to see that system 9 (the DINO training with a pretrained DINO model as the initial model) also achieves better a performance on Vox1 test set (not in VoxSRC21 Dev set) compared with the DINO baseline. But the relative less competitive EER comparing with the system 6-8 indicates that the contrastive learning based models do compensate for the non-contrastive training.  
\begin{table}[htbp]
\centering

\caption{\label{c3_eer} {Progressive results with the proposed techniques (EER in\% ), reported with the best ECAPA-TDNN model. }}
{\footnotesize{
\begin{tabular}{c|l|cc}\toprule
  system & method &  Vox1 test & VoxSRC21 Dev  \\
\cline{1-4}
 1 & MoCo   & 7.3  &  17.5 \\
 2 & + re-weighted infoNCE   & 6.9  &  16.9 \\
 3 & ++ProtoNCE (C3-MoCo)   & \textbf{6.4}  &  16.4 \\
 \midrule
 4 & DINO   & 3.3 &  12.7  \\
 5 & C3-DINO$_1$ (sys.3+4 multi-task)   & 3.4 &  13.2 \\
 6 & C3-DINO$_2$ (sys. 1 initial)   & 2.9 &  12.5 \\
 7 & C3-DINO$_2$ (sys. 2 initial)   & 2.7 &  12.4 \\
 8 & C3-DINO$_2$ (sys. 3 initial)   & \textbf{2.5} &  \textbf{12.2} \\
 9 & C3-DINO$_2$ (sys. 4 initial)   & 3.0 &  12.7 \\
\bottomrule
\end{tabular}}}
\vspace{-0.2cm}
\end{table}

\begin{figure}[htbp]
    \centering
    \includegraphics[width=0.85\linewidth, trim=4.5cm 9cm 4.5cm 8.3cm, height=5.7cm]{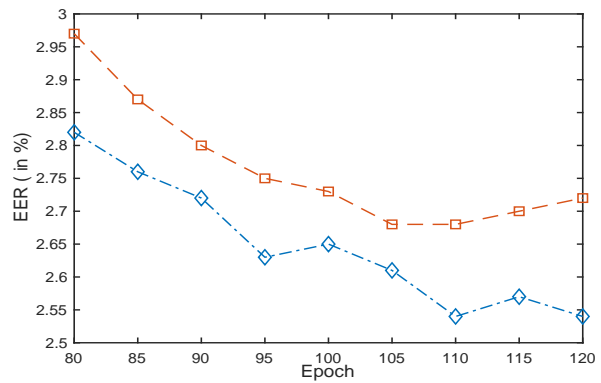}
    %\vspace{-ex}
    \caption{The difference between using the query/student encoder and using the key/teacher encoder for inference. Blue one is the key/teacher model.}
    \label{fig:teacher}
\end{figure}

\noindent \textbf{Key/teacher encoder for inference:} since there are two identical encoders in the SSL frameworks, we are also interested to see which one is better. \cref{fig:teacher} is the performance comparison between the query (student) and the key (teacher) encoder of the system 8. We can observe a clear performance advancement with the teacher encoder, which is consistent with the observation in~\cite{caron2021emerging}. It is noted that we did not find a similar trend in the MoCo based systems. Because of that reason, we only report the results from the teacher encoder for the DINO results in this study.  

\begin{figure}[htbp]
    \centering
    \includegraphics[width=0.85\linewidth, trim=5.5cm 8.8cm 5.5cm 9cm, height=6cm]{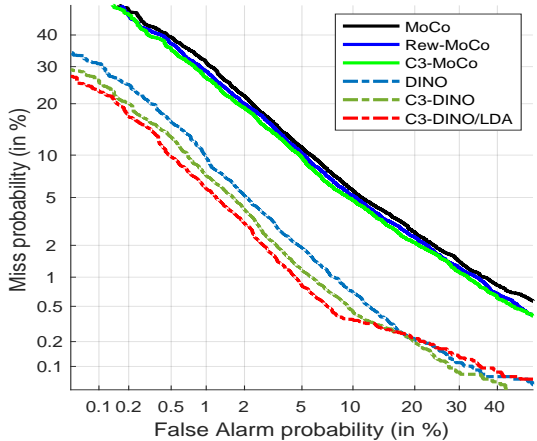}
    %\vspace{-ex}
    \caption{DET curves of the proposed SSL based SV systems on Vox1 test set.}
    \label{fig:det}
    
\end{figure}

\noindent \textbf{Self-training for the LDA/CDS back-end classifier:} to further improve the system performance, many previous studies rely on the methods that cluster the training set, assign the pseudo labels and retrain the speaker embedding system iteratively~\cite{desplanques2020ecapa,cho2021jhu}. However, this method has one major constrain: the clustering is based on the initial SSL speaker embeddings, to evaluate whether the clustering works or not, we have to use the whole data set with the pseudo labels to search for the optimal clustering results. That makes the process very expensive and difficult when the data set is in large scale with unknown speaker number. Since we have greatly improve the discriminative ability with the C3-DINO method. Our experiment on Voxceleb benchmark indicates a solution to further improve the SV task without iteratively training the speaker embedding system. By just performing the clustering over the C3-DINO speaker embeddings on Vox2 training data, training a LDA model to do the ``supervised" dimension reduction and applying the CDS classifier to get the final decision (clsuter number is 5000, dimension is reduced from 128 to 90), we are able to push the 2.5\% EER to 2.2\%, which is similar to the performance of the previous SOTA supervised DNN speaker embedding system before the advent of the ECAPA-TDNN model~\cite{chung2020defence}. The DET curve of the developed system are illustrated in \cref{fig:det}   

\subsection{Comprehensive system comparisons}
We list the recent development in self-supervised and semi-supervised speaker embedding systems for a comprehensive comparison. As shwon in \cref{sota3_eer}, we achieved the best contrastive learning based SV performance (C3-MoCo), we improved the SSL based SV performance by a big margin (2.5\% versus 4.8\%). Even compared with the classical supervised methods~\cite{snyder2018x}, the fully unsupervised models that we have built here are still competitive. 
\begin{table}[htbp]
\centering
\caption{\label{sota3_eer} { System comparisons on Vox1 test set (EER in\%). }}
{\footnotesize{
\begin{tabular}{c|l|c}\toprule
  supervision & method &  Vox1 test   \\
\cline{1-3}
  self-sup &  i-vector~\cite{huh2020augmentation} & 15.3   \\
 self-sup & AP+AAT~\cite{huh2020augmentation}   & 8.6   \\
 self-sup & MoCo+ProtoNCE~\cite{xia2021self}  & 8.2 \\
  self-sup & MoCo (E-TDNN)~\cite{thienpondt2020idlab} & 7.3 \\
  self-sup & Siaseme+SSR ~\cite{sang2021self} & 7.0 \\
  self-sup & C3-MoCo (system 3)   & \textbf{6.4}   \\
 \midrule
 self-sup & DINO~\cite{cho2021jhu}   & 4.8   \\
 self-sup & C3-DINO$_2$ (system 8)   & \textbf{2.5}  \\
 self-sup & C3-DINO$_2$ (system 8 + LDA)   & \textbf{2.2}  \\
 \midrule
semi-sup & GCL+PLDA (15\% label)~\cite{inoue2020semi}   & 6.0  \\
 semi-sup & MoCo+SupCon (15\% label)~\cite{xia2021self}   & 4.3  \\
  \midrule
 sup  & x-vector~\cite{povey2011kaldi}  & 3.1 \\
 sup & ECAPA-TDNN~\cite{desplanques2020ecapa}    & \textbf{0.9}  \\
\bottomrule
\end{tabular}}}
\vspace{-0.2cm}
\end{table}

\subsection{Performance across different scales of training data}
The last portion of this work is to evaluate how the SSL systems (baseline MoCo and DINO) evolve w.r.t. the different sizes of speaker number. For this experiment, we use the ``vehicle-spk" data, which can offer a relative big range of speaker numbers. Specifically, we use randomly picked 2k, 4k, 6k, 8k, 10k and 12k speakers to train the SSL models and a supervised contrastive model. The detailed EER results are listed in \cref{change_eer}. It is an empirical analysis. The general conclusion still holds the same as our Voxceleb studies. The DINO based models (the baseline DINO and the C3-DINO$_2$) and SupCon system outperforms the MoCo based systems (the baseline MoCo and the C3-MoCo). As demonstrated in the table, both non-contrastive and supervised models are having a good starting EER when the speaker number is small, while MoCo methods do not perform well in that condition. However, both DINOs and SupCon seem to be saturated when speaker number is large. While the MoCo systems are still improving with the increased speaker numbers. The experiment suggests that one of the conditions for contrastive SSL besed models to perform well is feeding more data to the training.   
\begin{table}[!h]
\caption{The SV performance (EER in \%) conditioned on different scales of training data.}
\label{change_eer}
\centering

\begin{tabular}{|c||c|c|c|c|c|c|}
\hline
method & 2k & 4k & 6k & 8k & 10k & 12k \\
\hline
MoCo  & 8.5 & 7.5 & 6.8 & 6.3& 5.7& 5.4 \\
C3-MoCo  & 8.1 & 7.2 & 6.6 & 5.9 & 5.3 & 4.9 \\
DINO  & 6.3 & 5.4 & 4.9 & 4.5 & 4.4 & 4.3 \\
C3-DINO$_2$  & 5.9 & 5.2 & 4.7 & 4.2 & 4.0 & 4.0 \\
SupCon~\cite{xia2021self}  & 5.2 & 4.6 & 4.2 & 3.9& 3.7 & 3.6 \\
\hline
\end{tabular}
\end{table}

\section{Conclusion}
\label{cons}
In this study, we investigated self-supervised learning based speaker embedding systems. By developing the MoCo and DINO baseline systems with different networks architectures and configurations, we were able to match and outperform the previous SSL based SV systems. To address the class-collision problem in the MoCo system, false negative filtering and infoNCE loss re-weighting were proposed based on the well designed experimental analysis. In combination with the ProtoNCE loss, the formulated C3-MoCo system achieved a 8.6\% relative improvement over the best contrastive learning based model~\cite{sang2021self}. And the proposed C3-DINO (with C3-MoCo assisted) speaker embedding system pushed the benchmark Vox2 test set to 2.5\% EER, a new record representing a 48.6\% relative improvement against the previous SOTA SSL based SV system. 

In addition to improve the SOTA of the SSL SV benchmarks. We also explored a simple and effective self-training method that further improved the result. The method show the potential to generalize well in a transfer learning scenario, where we just need to train an universal SSL model with all the available data, fine-tune a specific domain with in-domain data clustering, and train a back-end scoring classifier with the SSL speaker embeddings and pseudo labels generated from the clustering. We believe the greatly improved SSL based SV performance may serve as a trigger point to attract more attention for exploring the discrimination capability from the signal itself and reaching to a better generalization towards robust speaker recognition.

\section*{Acknowledgments}
The authors would like to thank Jeajin Cho from Johns Hopkins University for his many insights and helpful discussions during the development of this paper.

%\begin{thebibliography}{1}
\bibliographystyle{IEEEtran}
\bibliography{refs}

%\end{thebibliography}

%\newpage

\section{Biography Section}


\begin{IEEEbiography}[{\includegraphics[width=1in,height=1.25in,clip,keepaspectratio]{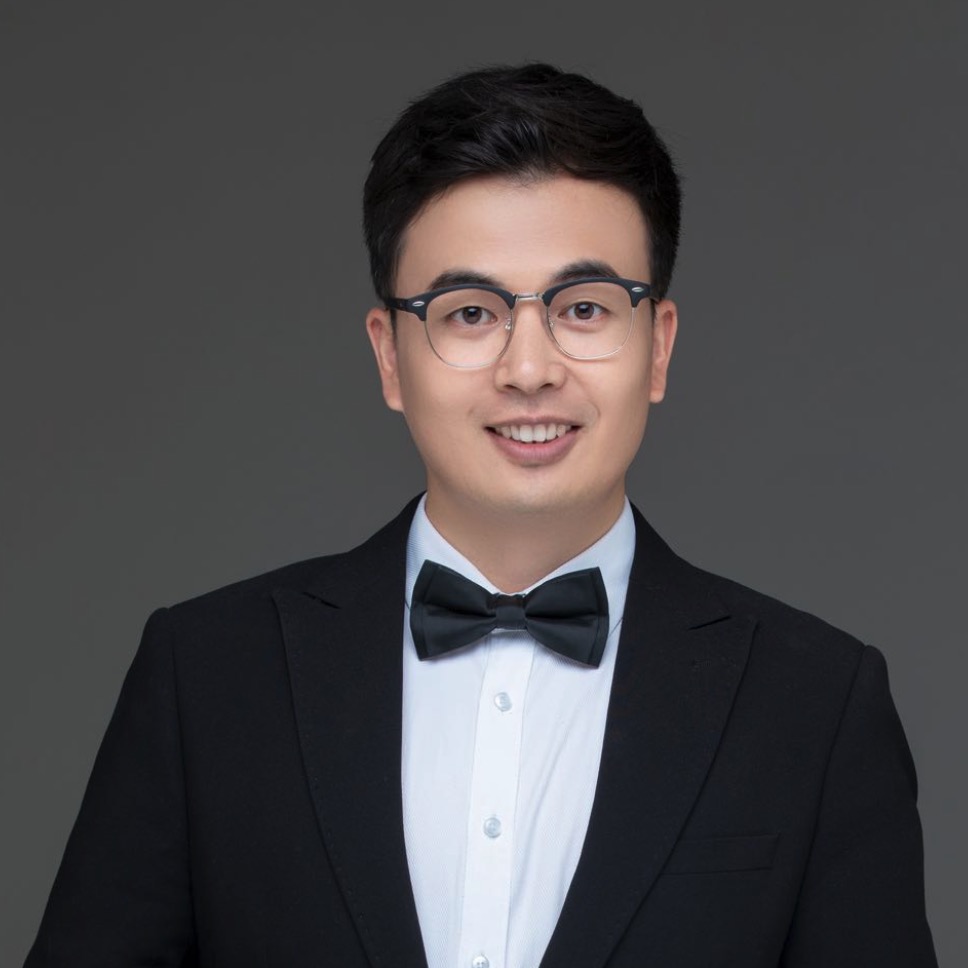}}]{Chunlei Zhang} (M'16) received the B.S. degree in environmental engineering and the M.S. degree in acoustics from Northwestern Polytechnical University, Xi’an, China, in 2011 and 2014, respectively. He received his Ph.D. degree in electrical and computer engineering from The University of Texas at Dallas, Richardson, TX, USA, in 2018. He joined the Tencent AI Lab, Bellevue, WA, USA, in 2019, where he is currently a Senior Research Scientist. His research interests include automatic speech recognition, speaker recognition/diarization and text-to-speech synthesis. He has been a member of the ISCA and IEEE since 2016. 
\end{IEEEbiography}

\begin{IEEEbiography}[{\includegraphics[width=1in,height=1.25in,clip,keepaspectratio]{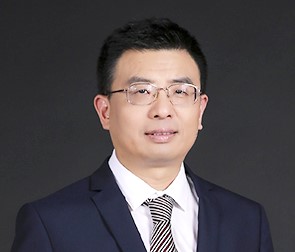}}]{Dong Yu} (M'97 SM'06 F'18) is an IEEE Fellow, an ISCA Fellow, and an ACM distinguished scientist. He currently works at Tencent AI Lab as a distinguished scientist and vice general manager. Prior to joining Tencent in 2017, he worked as a principal researcher at Microsoft Research (Redmond), where he had been since 1998. He has concentrated his research on speech recognition and processing, as well as natural language processing in recent years. He has two monographs and over 300 papers to his credit. His work has been widely cited, and he has received the prestigious IEEE Signal Processing Society best paper award in 2013, 2016, and 2020, as well as the 2021 NAACL best long paper award. He was a forerunner in the use of deep learning techniques in automatic speech recognition.

Dr. Dong Yu is currently serving as the chair of the IEEE Speech and Language Processing Technical Committee (SLTC). He has served on the editorial boards of numerous journals and magazines, as well as on the organizing and technical committees of numerous conferences and workshops, including serving as the technical co-chair of ICASSP 2021.
\end{IEEEbiography}

\vspace{11pt}

\vfill

\end{document}